\documentstyle[aps,epsf,preprint]{revtex}
\tightenlines
\begin{document}
\title{Quantum algorithms for Josephson networks}
\author{Jens Siewert$^{1,2}$ and Rosario Fazio$^{1}$}
\address{$^1$ Dipartimento di Metodologie Fisiche e Chimiche (DMFCI), 
	Universit\`a di Catania, viale A.Doria 6, I-95125 Catania, Italy\\
	Istituto Nazionale per la Fisica della Materia (INFM),
	Unit\`a di Catania, Italy\\
        $^2$ Institut f\"ur Theoretische Physik, Universit\"at Regensburg,
        D-93040 Regensburg, Germany}
\date{\today}
\maketitle
\begin{abstract}
        We analyze possible implementations of 
        quantum algorithms in a system of (macroscopic) 
        Josephson charge qubits. System layout and parameters to
        realize the Deutsch algorithm with up to three qubits
        are provided. Special attention is paid to the 
        necessity of entangled states in the various implementations.
        Further, we demonstrate
        explicitely that the gates to implement 
        the Bernstein-Vazirani algorithm
        can be realized by using a system of uncoupled qubits.
\end{abstract}
Quantum information processing has initiated an
impressive research activity throughout the scientific 
community during the past decade.
The interest in quantum computation, in particular, is stimulated 
by the discovery of quantum algorithms~\cite{Shor94,Grover97} which 
can outperform their classical counterparts in solving problems of 
significant practical relevance.

In recent years, considerable progress has been made in the field
of ``quantum hardware'', 
{\it i.e.\/} in the search for physical systems appropriate for the
implementation of quantum computation. 
It is equally important in the developement of quantum computing
to experimentally realize complete quantum algorithms.
To date, this has been achieved in liquid-state NMR
\cite{Gershenfeld97,Cory97,Chuangetal98,JonesMoscaHansen,Linden98,%
      Arvind00,Arvind99,Collins00}, 
in atomic physics \cite{Ahn2000} and by
optical interferometry \cite{Kwiat99}.
A solid-state implementation
of Grover's algorithm has recently been proposed in Ref.~\cite{Loss2000}.

The quest for large scale integrability has stimulated 
an increasing interest in  superconducting 
nanocircuits \cite{ShnirmanPRL,Averin98,Makhlin99,Caspar99,nature}
as possible candidates for the implementation of a quantum computer.
The recent experimental breakthrough for Josephson 
qubits~\cite{Nakamura99,Friedman,Caspar00} is the first important step 
towards a solid-state quantum computer.

Naturally the question arises whether, at the present
technological level, it is possible 
to implement also quantum algorithms in these systems. 
Here we concentrate on charge 
qubits~\cite{ShnirmanPRL,Averin98,Makhlin99} and show how the 
Deutsch algorithm \cite{Deutsch85,Deutsch92,Cleve98,Collins98} and the 
Bernstein-Vazirani algorithm \cite{Bernstein93} can be run on a 
Josephson quantum computer. 
We analyze the experiment by Nakamura {\it et al.\/} \cite{Nakamura99}
in terms of quantum interferometry \cite{Cleve98} and show 
that it corresponds to the implementation
of the one-qubit version
of Deutsch's algorithm. By generalizing this idea we show how 
the $N$-qubit Deutsch algorithm, with $N\leq3$, can be implemented. 
Entanglement is required only for
$N\geq3$ \cite{Collins98}. Finally we show
explicitely that the Bernstein-Vazirani algorithm~\cite{Bernstein93} 
can be implemented using  uncoupled qubits (for arbitrary $N$). Therefore 
it can be realized by means of the setup of Ref.\cite{Nakamura99}.

Consider the subset of Boolean functions $f:\ \{0,1\}^N \rightarrow \{0,1\}$ 
with the property that $f$ is either constant or balanced (that is,
it has an equal number of 0 outputs as 1s). 
The Deutsch-Jozsa 
algorithm \cite{Deutsch85,Deutsch92,Cleve98} determines 
whether a function $f$ is constant or balanced. 
With a classical algorithm, this problem would, in
the worst case, require $2^{N-1}+1$ evaluations of $f$ whereas the
quantum algorithm solves it with a single evaluation by means of the following 
steps (here we focus on the  refined version 
by Collins {\it et al.\/}~\cite{Collins98},
see also Fig.\ 1). 
\newline\indent
{\it (i)} All qubits are prepared in the initial state $|0\rangle$,
therefore the $N$-qubit register is in the state $|00\ldots 0\rangle$.
\newline\indent
{\it (ii)} Perform an $N$-qubit Hadamard  
transformation $\cal H$ 
\begin{equation}
|x\rangle \ \stackrel{\cal H}{\longrightarrow} \sum_{y\in \{0,1\}^N}
                                             (-1)^{x\cdot y}\ |y\rangle\ , 
           \ \ \ (x\in \{0,1\}^N),
\label{HadamardTrafo}
\nonumber
\end{equation}
where
$
     x\cdot y = (x_1 \wedge y_1) \oplus \ldots \oplus (x_N \wedge y_N)
$
is the scalar product modulo two. This is equivalent to 
performing a one-bit Hadamard transformation to each qubit individually.
\newline\indent
{\it (iii)} Apply the $f$-controlled phase shift $U_f$ 
\cite{Cleve98,Collins98}
\begin{equation}
    |x\rangle\ \stackrel{U_f}{\longrightarrow}\ (-1)^{f(x)}\ |x\rangle\ ,\ \ \
        (x\in \{0,1\}^N)\ .
\label{f-phaseshift}
\end{equation}
 Note that we will use the convention $f(00\ldots 0)=0$. 
\newline\indent
{\it (iv)} Perform another Hadamard transformation $\cal H$. 
\newline\indent
{\it (v)} Measure the final state of the register. If the result
is $|00\ldots 0\rangle$ the function $f$ is constant; if, however,
 the amplitude $a_{|00\ldots0\rangle}$ of the state 
$|00\ldots 0\rangle$ is zero the function $f$ is balanced.
This is because 
\begin{equation}
     a_{|00\ldots 0\rangle} = \frac{1}{2^N}
                                 \sum_{x\in \{0,1\}^N} (-1)^{f(x)}\ \ .
\label{amplitude0}
\end{equation}
Entanglement is not needed for the one-bit and two-bit case~\cite{Collins98}
while it is necessary for the exponential speedup for a higher
number of qubits~\cite{Azuma01}.  

In order to implement the algorithm we have to show that each
individual step (preparation of the state, gate operations, measurement)
can be realized. It is well known how to prepare
and to measure the states in Josephson charge qubits 
\cite{ShnirmanPRL,Makhlin99,Nakamura99}. Our task is to demonstrate
that the gate operations corresponding to {\it all} possible functions
$f$ can be performed with a single device. An important aspect of our
proposal is that the gate operations are represented in a basis of
superpositions of charge states.

{\it One-qubit and two-qubit Deutsch algorithm -\ } 
In the one-bit case 
there is one \mbox{constant} function and one balanced function $f$
(due to our choice $f(0)=0$, see above).
The gate $U_f$ implementing the constant function is 
the one-qubit identity operator $I_1$. 
The balanced function can be represented (with respect to
the computational basis) by $\sigma_z$
where $\sigma_i$ denote the Pauli matrices.
%

The sequence of steps {\it (i)-(v)} 
can be carried out with a Josephson qubit.
A Josephson charge qubit \cite{ShnirmanPRL,Makhlin99}
is a Cooper-pair box (see Fig.\ 3a) 
which is characterized by two energy scales, the charging energy 
$E_{\rm ch}=(2e)^2/(2C)$ (here $C$ is the total capacitance of
the island) and the Josephson energy $E_J\ll E_{\rm ch}$ 
of the tunnel junction. 
At low temperatures $T\ll E_J$ 
only the two charges states with 0 and 1 excess
Cooper pair on the island are important and 
the system behaves as a two-level system with the Hilbert space 
$\{|0\rangle,|1\rangle \}$ and the one-qubit Hamiltonian
\begin{equation}
    H_{\rm 1q}= (E_{\rm ch}/2)\ (2n_x-1)\ \sigma_z 
                           \    -\ (E_J/2)\ \sigma_x \ \ .
\label{1qHamiltonian}
\end{equation}
Here $n_x=C_g V_g/(2e)$ is the offset charge which can 
be controlled by the gate voltage. 

The one-bit version of the Deutsch algorithm is already realized 
in the experiment by Nakamura {\em et al} (see Figs.\ 1 and 2).
%
%
%
First the system is prepared in a symmetric superposition of the states.
The Rabi oscillation in the experiment corresponds to the action of the 
controlled phase shift $U_f$. Finally the system is measured.
Note, however, an important difference between the experiment and 
the steps {\it (i) - (v)}: 
In the usual representation of  Deutsch's algorithm 
the gate $U_f$ 
acts on the same basis states which then are measured;
the Hadamard transformation produces the appropriate superpositions. 
In contrast to this, the measurement in the experiment is done
in the charge basis while the ``gate'' acts in the
basis $\{|+\rangle,|-\rangle\}$ 
which is related to the charge basis by a Hadamard transformation.
By identifying $U_f \leftrightarrow \exp{(i(E_Jt/2\hbar)\sigma_z)} $ the 
different functions $f$ can be implemented by choosing the time $t$
appropriately (see Table I).
%
%
%

This observation suggests the possibility to implement  the Deutsch algorithm 
in a setup of more than one charge qubit by performing the same 
sequence of gate voltage pulses
as in the experiment of Ref.\ \cite{Nakamura99}. 
The gates should operate at the degeneracy 
point $n_x^{(j)}=1/2$ of the charge qubits. What we need is to find the 
proper parameters and operation times to obtain all 
possible gates $U_f$. 

For two qubits this realization is obvious as
the two-qubit algorithm can be 
implemented by using two uncoupled qubits \cite{Collins98}. 
The gates 
$\sigma_z^{(1)} \otimes I_1^{(2)}$, $I_1^{(1)} \otimes \sigma_z^{(2)}$,
$\sigma_z^{(1)}\otimes \sigma_z^{(2)}$ implementing the balanced functions
(the upper index denotes the qubit number) and
$I_1^{(1)}\otimes I_1^{(2)}$ for the constant function can be realized
in complete analogy to the one-qubit algorithm.

{\it Three-qubit Deutsch algorithm -\ } 
The realization of the three-qubit version of the algorithm is 
more difficult. 
Apart from the constant function 
35 balanced functions need to be implemented. Moreover,
the three-qubit algorithm involves gates $U_f$ which produce
entangled final states. 

The goal is to proceed along the same lines as above, that is, 
preparation of the state
$|000\rangle$, sudden sweep of $n_x^{(j)}$ etc.
The action of the gates $U_f$ takes place
in the basis \mbox{$\{|+++\rangle$},$|++-\rangle,\ldots,|---\rangle\}$.
In order to find 
efficient ways for the implementation we first analyze the
functions $f$ and the corresponding gates $U_f$.

Apart from the constant function and its gate 
$I^{(1)}_1\otimes I^{(2)}_1\otimes I^{(3)}_1$ there are 7 balanced
functions for which the gates are separable: 
$\sigma^{(1)}_z \otimes I^{(2)}_1\otimes I^{(3)}_1$,
$I^{(1)}_1\otimes \sigma^{(2)}_z\otimes I^{(3)}_1$, 
\ldots,
$\sigma_z^{(1)}\otimes \sigma_z^{(2)}\otimes \sigma_z^{(3)}$.
Further there are 12 balanced functions for which the 
gates factorize into a one-qubit part and a two-qubit
part as in example $I)$ below. 
The other gates entangle all three qubits and can be divided
into two classes (see example $II)$ and $III)$).
There are 12 gates of class $II)$ and 4 gates of class $III)$.
\[
\begin{array}{ll}
I) &
            \displaystyle\frac{1}{2}\
     \left(    I^{(1)}_1\otimes I^{(2)}_1
             + \sigma^{(1)}_z\otimes I^{(2)}_1   
            - I^{(1)}_1\otimes \sigma^{(2)}_z
            + \sigma^{(1)}_z\otimes \sigma^{(2)}_z
     \right)
              \otimes \sigma_z^{(3)}
\\
II)  &
            \displaystyle\frac{1}{2}
            \left( \sigma_z^{(1)} \otimes I^{(2)}_1\otimes I^{(3)}_1 
        -  I^{(1)}_1\otimes I^{(2)}_1 \otimes \sigma_z^{(3)} 
                +  \sigma_z^{(1)} \otimes \sigma^{(2)}_z\otimes I^{(3)}_1
            +   I^{(1)}_1\otimes \sigma^{(2)}_z \otimes \sigma_z^{(3)}
            \right)  
 \\
III) &
            \displaystyle\frac{1}{2}
            \left( \sigma_z^{(1)} \otimes I^{(2)}_1\otimes I^{(3)}_1 
        -  I^{(1)}_1\otimes \sigma^{(2)}_z \otimes I_1^{(3)} 
                +  I_1^{(1)} \otimes I^{(2)}_1\otimes \sigma^{(3)}_z
            +   \sigma^{(1)}_z\otimes \sigma^{(2)}_z \otimes \sigma_z^{(3)}
            \right)
\end{array}
\]
All separable qubit operations can be carried out in analogy 
with the one-qubit case above. In the following we discuss
how the entangling gate operations can be achieved.
For the realization of these gates a coupling of tunable strength
between the qubits is required. 

There are 
various ways to couple charge qubits 
\cite{ShnirmanPRL,Averin98,Makhlin99,nature}. 
Here we investigate 
coupling via
Josephson junctions \cite{ourJLTP}. 
Each qubit island is coupled to its nearest neighbor
using a symmetric SQUID
(see Fig.\ 3b). 

Assuming that both the $j$-th qubit and the $j$-th
coupling junction are tunable by local fluxes $\Phi^{(j)}$, $\Phi^{(j)}_K$
the Hamiltonian for the $N$-qubit system at 
the degeneracy point $n_x^{(j)}=1/2$ reads
\begin{equation}
  \begin{array}{cl}
    H_{N{\rm q}} & = \displaystyle
                     \sum_{j=1}^N \ \ 
                                  \left\{ 
                                 H_{\rm 1q}^{(j)}(\Phi^{(j)}) 
   \    +\   E_K^{(j)}\  \sigma_z^{(j)}\sigma_z^{(j+1)}
                                  \right.
              \\[4mm] &           \left.
              - (1/2)J_K^{(j)}(\Phi_K^{(j)})\ 
                 [\ \sigma^{(j)}_+ \sigma^{(j+1)}_- \ +\ \ {\rm h.c.\/}\ ]
                                \right\}\ \ .
  \end{array}
\label{NqHamiltonian}
\end{equation}
Here $J_K^{(j)}$ is the Josephson energy of the $j$-th coupling SQUID
and $\sigma_\pm=(\sigma_x\pm i\sigma_y)/2$. For small coupling capacitance
$C_K^{(j)}\ll C^{(j)}$ we have
$E_K^{(j)}=(C_K^{(j)}/C^{(j)})E_{\rm ch}^{(j)}/2$. 
We will assume that $E_K^{(j)}$ is negligible.
Since in practice the capacitive coupling is
always present it is necessary to have $J_K^{(j)}(\Phi=0)\gg 4E_K^{(j)}$.
Then the dynamics of the system approximates the ideal dynamics 
sufficiently well. 

Consider now the first and the second qubit coupled by $J_K^{(1)}$.
By choosing, {\it e.g.,} 
$-E_J^{(1)}=E_J^{(2)}=\pm J_K^{(1)}={\sf J}$ and the operation
time $t\simeq0.97 (2\pi/{\sf J})$ we obtain 
an operation similar to a swap gate for the qubits 1 and 2
for which we introduce the notation
(in the basis $\{|++\rangle,\ldots,
                   |--\rangle \}_{12}$) 
\begin{equation}
        [\pm 12] :=: \left(
                              \begin{array}{cccc}
                                0 & 0 & 0 & \pm i \\
                                0 & 1 & 0 & 0 \\
                                0 & 0 & 1 & 0 \\
                                \pm i & 0 & 0 & 0 
                              \end{array}
                   \right) \ \ .
\end{equation}
By denoting one-bit phase shifts for the $j$-th qubit 
\begin{equation}
        [\pm j] :=: \left(
                              \begin{array}{cc}
                                1 & 0 \\
                                0 & \pm i 
                              \end{array}
                   \right) \ \ ,
\end{equation}
we can write a sequence of operations which gives the two-bit
entangling gate in example $I)$ above:
\begin{equation}
                       [+1][-2] \frac{\rule{.4cm}{.0cm}}{\rule{.4cm}{.0cm}}
                       [-12] 
   \frac{\rule{.4cm}{.0cm}}{\rule{.4cm}{.0cm}}   \sigma^{(3)}_z   
                       \ \ .
\label{exI}
\end{equation}
After suddenly sweeping $n_x^{(1)}$ and $n_x^{(2)}$ to the degeneracy,
first the one-bit phase shifts are performed while $J_K^{(1)}=0$.
Then $J_K^{(1)}$ is switched on suddenly in order to do the
two-bit rotation. The $\sigma^{(3)}_z$ rotation can be done at any moment
since the third qubit is decoupled from the other two. Finally
the $n_x^{(j)}$ are swept back suddenly and the register is measured.

There are numerous ways to represent the three-bit entangling gates. 
At least two different two-bit rotations need to be applied.
During the second two-bit rotation the third qubit has to be ``halted''.
This can be done by switching off both the $E_J$ and the $J_K$ which
couple to this qubit. 
A possible sequence for example $II)$ is
\begin{equation}
                       [+1][-2] \frac{\rule{.4cm}{.0cm}}{\rule{.4cm}{.0cm}}
                       [+13] \frac{\rule{.4cm}{.0cm}}{\rule{.4cm}{.0cm}}
                       [-12] 
                       \ \ ,
\label{exII}
\end{equation}
and for example $III)$
\begin{equation}
 \sigma_z^{(1)}\sigma_z^{(2)}\sigma_z^{(3)}
                       \frac{\rule{.4cm}{.0cm}}{\rule{.4cm}{.0cm}}
 [+12]                 \frac{\rule{.4cm}{.0cm}}{\rule{.4cm}{.0cm}}
 [+23]                 \frac{\rule{.4cm}{.0cm}}{\rule{.4cm}{.0cm}}
 [+12] \ \ .
\label{exIII}
\end{equation}
 The complete set of entangling gates can be obtained from the sequences 
 (\ref{exI}) - (\ref{exIII})
 by cyclic permutations of qubit numbers (and appropriate sign changes),
 thereby paying attention that the parameter settings are compatible for
 both one-bit and two-bit operations.
 We note that universality in quantum computation implies that given 
 {\it some}
 two-bit gate, it is possible to realize any algorithm. 
 This procedure, however,
 in general requires much longer sequences of one-bit and two-bit gates.
 Our proposal allows the implementation of simple algorithms following
 the scheme given in Fig.\ 1 and is amenable of an experimental verification
 with present-day technology.
It is interesting to note that the completely entangling gates 
of class $II)$  and $III)$ can be realized approximately
with a {\it single} three-qubit operation.
 In Table II we list the parameters for the various implementations
 including estimates for the accuracy of the respective operation.

{\it The Bernstein-Vazirani algorithm} \cite{Bernstein93,Cleve98} -\ 
It is analogous to the Deutsch algorithm described
in the beginning, with the difference that 
the function $f$ has the form 
$ f = a\cdot x\oplus b$, $(a,x\in \{0,1\}^N,b\in \{0,1\})$.
By measuring the register after running the algorithm once
(the gate in step {\em (iii)} is denoted by $U_a$)
one obtains the number $a$ in binary representation which
classically would require $N$ function calls.
The fact that there is no entanglement in the Bernstein-Vazirani
algorithm has been observed in Ref.\ \cite{Meyer}. 
Here we demonstrate explicitely that the algorithm can be obtained
by applying only one-qubit operations.

The gate $U_a$ can be rewritten as a product of single-qubit
gates (using the definition $(\sigma_k)^0 :=: I_1$)
\begin{equation} 
 U_a = (-1)^b \ \prod_{j=1}^N (\sigma_z^{(j)})^{a_j}
\end{equation} 
where $a_j$ denotes the $j$-th digit of $a$ in binary representation.
Apart from the global phase $(-1)^b$ this is the part of the Deutsch
algorithm with completely separable gates. As the algorithm starts with a 
product state, no entanglement is involved at any step. 
(We note that one can rewrite the action of the 
whole algorithm ${\cal H}U_a{\cal H}$
as $\prod_j (\sigma_x^{(j)})^{a_j}$ which trivially gives the result.)
It is therefore possible to realize the Bernstein-Vazirani algorithm 
with Josephson networks in complete analogy with the implementation
for the one-qubit and two-qubit Deutsch algorithm.

\acknowledgements
The authors would like to thank L.\ Amico, D.V.\ Averin,  I.\ Chuang, 
P.\ Delsing, G.\ Falci, J.B.\ Majer,
Y.\ Makhlin, A.\ Osterloh, G.M.\ Palma, F.\ Plastina, 
C.\ Urbina, V.\ Vedral and C.\ v.d.\ Wal 
for stimulating discussions. 
This work was supported in part by the EC-TMR, IST-Squbit
and INFM-PRA-SSQI.

\begin{table} 
\begin{tabular}{|c|c|c|}\hline
     f     & \parbox{7cm}{  \begin{center}
                              gate $U_f$
                            \end{center}  }  
           & \parbox{7cm}{  \begin{center}
                              time $t$ 
                            \end{center}  }     \\ \hline
     constant & $I_1$         & $2\pi\hbar/E_J$ \\ \hline
     balanced & $\sigma_z$    & $\pi\hbar/E_J$ \\ \hline 
\end{tabular} 
\caption{}
\end{table} 
\vspace*{3cm}
\begin{table}
\begin{tabular}{|c|c|c|c|c|c|c|c|c|c|c|}\hline
  gate & implementation & $E_J^{(1)}$ & $E_J^{(2)}$ & $E_J^{(3)}$    
                        & $J_K^{(1)}$ & $J_K^{(2)}$ & $J_K^{(3)}$    
                        & time $t/(2\pi\hbar/{\sf J})$ 
                        & \parbox{2.cm}{\begin{center}
                                             $a_{|0\ldots0\rangle}$\\
                                             $(E_K^{(j)}=0)$
                                        \end{center}} 
                        & \parbox{2.3cm}{\begin{center}
                                             $a_{|0\ldots0\rangle}$\\
                                             $(E_K^{(j)}={\sf J}/40)$
                                        \end{center}} 
                        \\ \hline\hline
 $II)$ & sequence (9)   & {\sf -J} & {\sf J} & {\sf J} 
                        & {\sf -J} &  0      & {\sf J} 
                        & 0.97 (2bit op.\/)
                        &  2 $\cdot$ $10^{-3}$     & 2 $\cdot$ $10^{-4}$ 
                        \\ \hline
 $II)$ & single operation   & {\sf -J/2} & 0       & {\sf J}/2 
                        & {\sf J} & {\sf J} & 0
                        & 0.80
                        & 7 $\cdot$ $10^{-5}$     &  2 $\cdot$ $10^{-2}$
                        \\ \hline
$III)$ & sequence (10)  & {\sf J} & {\sf -J}& {\sf J} 
                        & {\sf J} & {\sf J} & 0 
                        & 0.97 (2bit op.\/)
                        & 3 $\cdot$ $10^{-4}$     &  6 $\cdot$ $10^{-3}$
                        \\ \hline
$III)$ & single operation   & {\sf J}/2 & {\sf -J}/2 & $0.83 {\sf J}$ 
                        & {\sf 0} & {\sf J} & {\sf J} 
                        & 1.19
                        & $< 10^{-5}$     &  2 $\cdot$ $10^{-3}$
                        \\ \hline
\end{tabular} 
\caption{Parameters for various realizations of the gates $II)$ and $III)$.
         The coefficient $a_{|0\ldots0\rangle}$ is a measure for the 
         fidelity of the operation (for an ideal operation
         $\ a_{|0\ldots0\rangle} = 0$). The operation time for the sequences
         refers to the time needed for the two-qubit rotations.
         Single-qubit rotations are assumed to be perfect.}
\end{table}  

\newpage
\begin{figure}
\vspace*{1cm}
\caption{The sequence of operations to perform the Deutsch algorithm 
         on a register of $N$ qubits. According to Ref.\
         \protect\cite{Cleve98} it  can be interpreted in terms 
         of quantum interferometry. The first Hadamard transformation
         produces a superposition of {\it all} possible states.
         Thus, with the application of the $f$-controlled gate $U_f$
         the outcome of $f$ for all possible arguments is evaluated
         at the same time.
         The second Hadamard transformation brings all computational
         paths together.} 
\end{figure}
\begin{figure}
\vspace{2cm}
\caption{Schematic representation of Nakamura's experiment 
           \protect\cite{Nakamura99}.
         The qubit is prepared in the ground state $|0\rangle$.
         After suddenly sweeping the gate voltage the system starts 
         Rabi oscillations between the eigenstates of the new
         Hamiltonian \protect$|\pm\rangle = (|0\rangle\pm |1\rangle)/\sqrt{2}$.
         After the time $t$ the gate voltage is swept back suddenly
           which freezes the final state; then the qubit is measured.} 
\end{figure}

\begin{figure}
\vspace{2cm}
\caption{a) A charge qubit. The Josephson energy of the junction
         can be controlled by the magnetic flux $\Phi$:
         $E_J(\Phi)=2{\cal E}_J \cos{(\pi\Phi/\Phi_0)}$, where
         ${\cal E}_J$ is the Josephson energy of the junctions
         of the symmetric SQUID and $\Phi_0=h/(2e)$
         \protect\cite{Makhlin99}.
         Typical parameters are ${\cal E}_{J}\sim 30\mu$eV
         and $E_{\rm ch}\sim 500\mu$eV.
         b) A possible realization of coupled charge qubits.}
\end{figure}

\newpage
\begin{figure}
\vspace*{3cm}
\centerline{{\epsfxsize=16.5cm\epsfysize=3.5cm\epsfbox{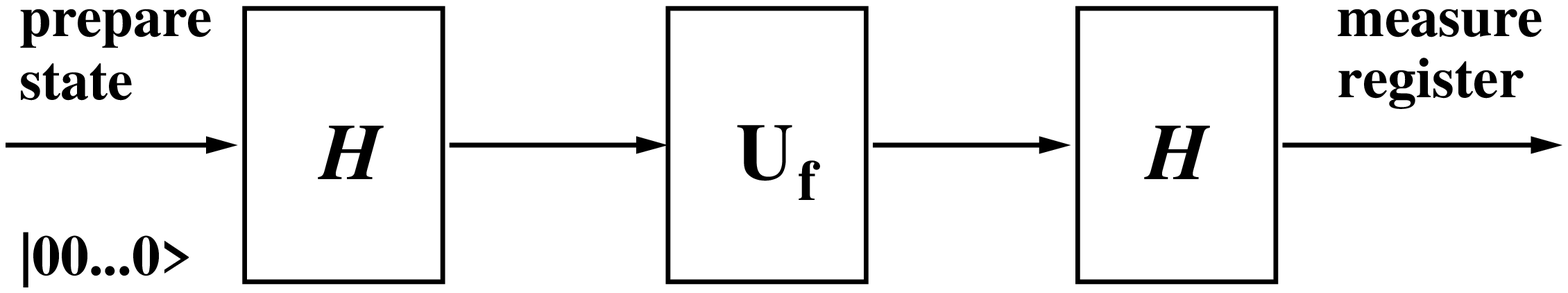}}}
\vspace{.5cm}
\end{figure}

\newpage
\begin{figure}
\vspace*{3cm}
\centerline{{\epsfxsize=16.5cm\epsfysize=3.5cm\epsfbox{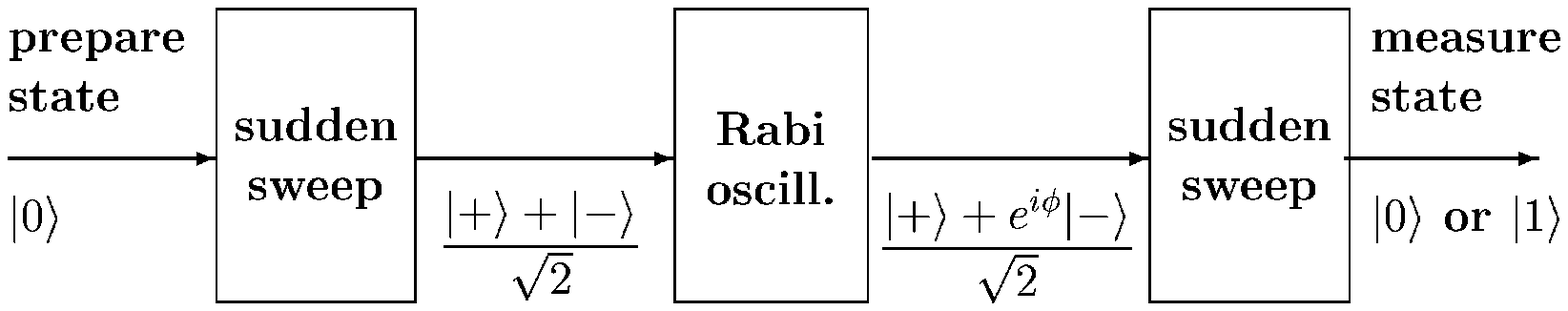}}}
\vspace{.5cm}
\end{figure}

\newpage
\begin{figure}
\vspace*{2cm}
\centerline{{\epsfxsize=16.cm\epsfysize=7.5cm\epsfbox{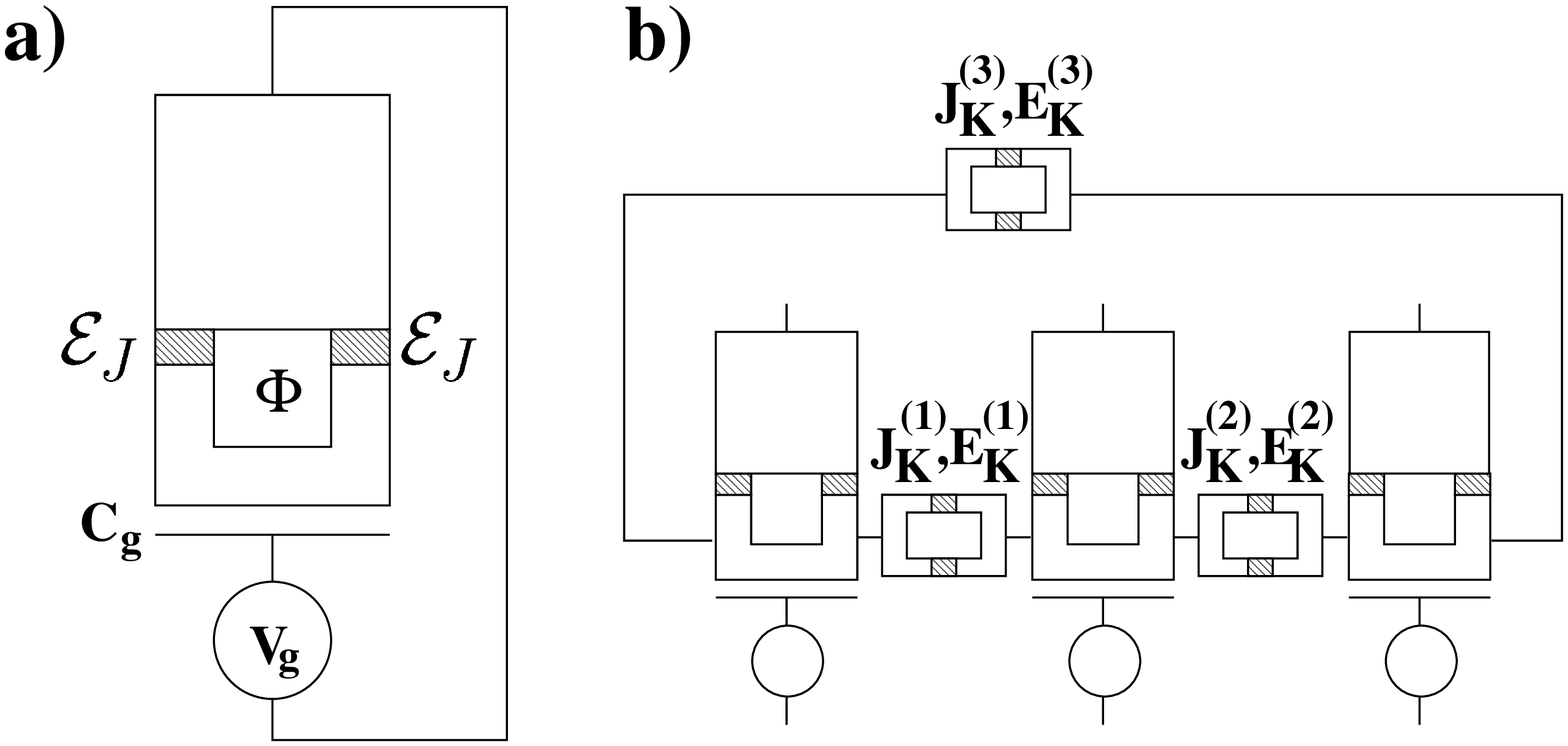}}}
\end{figure}
\end{document}